\documentclass[fleqn,11pt]{wlscirep}

\usepackage[utf8]{inputenc}
\usepackage[T1]{fontenc}
\usepackage{todonotes}
\usepackage{verbatim}
\usepackage{microtype}
\usepackage{geometry}
\title{Preparing for the Future -- Rethinking Proxy Apps}

\author[1]{Satoshi Matsuoka}
\author[1]{Jens Domke}
\author[1]{Mohamed Wahib}
\author[1]{Aleksandr Drozd}
\author[2]{Ray Bair}
\author[2]{Andrew A. Chien}
\author[3]{Jeffrey S. Vetter}
\author[4]{John Shalf}

\affil[1]{RIKEN Center for Computational Science, Kobe, 650-0047, Japan}
\affil[2]{Argonne National Laboratory, Lemont, IL 60439, USA}
\affil[3]{Oak Ridge National Laboratory, Oak Ridge, TN 37831-6173, USA}
\affil[4]{Lawrence Berkeley National Laboratory, Berkeley, CA 94720-8150, USA}

\begin{abstract}
A considerable amount of research and engineering went into designing proxy applications, which represent common high-performance computing workloads, to co-design and evaluate the current generation of supercomputers, e.g., RIKEN's Supercomputer Fugaku, ANL's Aurora, or ORNL's Frontier. 
This process was necessary to standardize the procurement while avoiding duplicated effort at each HPC center
to develop their own benchmarks. Unfortunately, proxy applications force HPC centers and providers (vendors) into a an undesirable state of rigidity, in contrast to the fast-moving trends of current technology and future heterogeneity. To accommodate an extremely-heterogeneous future, we have to reconsider
how to co-design supercomputers during the next decade, and avoid repeating the past mistakes. 
This position paper outlines the current state-of-the-art in system co-design, challenges encountered over the past years, and a proposed plan to move forward.
\end{abstract}

\begin{document}

\flushbottom
\maketitle
\thispagestyle{empty}

\section{Introduction}
Supercomputing is the art of mapping a scientific question onto hundreds of trillions or
quadrillions of transistors, as in the case of the currently fastest supercomputers in the world, by exploiting the
problem's underlying concurrency. Unfortunately, this requires numerous transformations:
\textit{question->algorithm->parallelization->language->compilation->execution},
and intermediate bottlenecks,
such as Amdahl's law, are complicating an efficient utilization of the available transistors. While society's
problems are somewhat immutable, until solved, we see an increase in available choices in the remainder of
this chain of transformations.

Historically, the sciences' demand for more compute capabilities was met by increasing the transistor count and
density, and hence the users had autonomy all the way from problem to compilation, while the HPC community
focused primarily on developing parallelization techniques and focused on perfecting component integration to
assemble the supercomputers. But the projected end of Moore's law and Dennard's scaling in the early 2000s
required a rethinking, culminating in an intensified co-design effort at the HPC centers around the world.
This meant that supercomputing centers had to take a closer look at the workloads, executed by its users,
to design the best architecture that meets the computational demands. The resulting scaled-down versions
of important scientific applications, so-called \textit{mini} or \textit{proxy applications}~\cite{role_of_codesign},
which represent the workload from problem to language, redefined a new overlapping between HPC users, centers,
and vendors. HPC centers and vendors tailored the hardware architectures, i.e., many-core CPUs and/or GPUs paired with high-bandwidth
memory, and improved the languages, such as CUDA, to meet the proxy's requirements; while the users were
required to modify their parallelization approaches and data layout, and potentially had to rewrite certain
kernels in other languages.

We believe that this approach, while being somewhat successful for the current generation of (pre-)exascale
supercomputers, is not sustainable in the future. The primary reasons being an explosion in workload
complexity, number of proxy applications, hardware choices, programming languages, and parallelization strategies.
All while the amount of financial and human resources to tackle the problem remains finite and limited.
For example, important scientific applications such as climate simulations have evolved into multi-million
lines of code comprising many algorithms (and their respective kernels) which either execute in parallel
or sequentially (see coupled climate models~\cite{cesm}) depending on availability of computational resources
and depending on the inherent load imbalance of the underlying problem (e.g., the location of clouds in the simulated area).
Additionally, these workloads may include extensive pre-/post-processing of the data streams or they may
assimilate new data from external sensors on the fly~\cite{9355206}. The availability of Noisy Intermediate-Scale
Quantum (NISQ) computers, FPGAs and ASICs, in-memory processing, SmartNICs, and the commercial success of Deep Learning
and crypto-currencies, increases the choice for processors beyond just traditional CPUs and GPGPUs (see Cambrian explosion in processor
architecture~\cite{10.1145/3208040.3225055}).
Furthermore, domain experts strive for higher-level languages, such as Julia
or Python, or domain-specific languages (e.g., Tensorflow) to increase productivity; while HPC experts
seek performance portability and offer new ways to exploit concurrency with OneAPI, Kokkos, RAJA, Chapel, etc.

Hence, the intersection between HPC users and providers has to change, yet again, to improve future supercomputing.
The questions we want to address in the upcoming sections is: What is a better alternative to current
proxy applications to tackle this increasing complexity? Is it possible to maintain an efficient mapping from problem to
execution within the given limits? Can we improve, automate, and streamline the co-design and procurement process
for the next generation of supercomputers?

\section{State-of-the-Art for Co-Design Tool Chains and Procurement Benchmarks}
To aid the co-design in recent years, governments around the world have tasked their HPC centers with developing
a set of benchmarks that reflect the priority applications of each individual nation. Those benchmarks would be used
to drive the development process of the current generation of (pre-)exascale systems, which either came online
recently or are scheduled to be operational in the coming months. The outcome of those efforts are, for example, the
Exascale Computing Project (ECP) Proxy Applications~\cite{exascale_computing_project_ecp_2018} (14 applications),
Center for Efficient Exascale Discretizations (CEED) Miniapps~\cite{exascale_computing_project_ceed_2020} (6 codes) and
CORAL-2 Benchmarks~\cite{llnl_coral2_nodate} (21 benchmarks or benchmark sets), all developed by the DoE national labs of the USA.
The European Union's PRACE consortium hosts the Unified European Application Benchmark Suite (UEABS)~\cite{prace_unified_2016}
with 13 application codes reflecting the sciences performed in the EU. Furthermore, RIKEN assembled a list of
8 Fiber/proxy applications~\cite{riken_aics_fiber_2015}, spanning the social and scientific
priority areas of Japan, to assist in the development of hardware and software for the Supercomputer Fugaku~\cite{sato_co-design_2020}.

In the case of Fugaku, the Fiber applications were used by researchers at RIKEN and Fujitsu, among many things,
to analyze performance requirements from application metrics such as bytes-to-flop ratio and to develop new metrics,
e.g. Simplified Sustained System Performance (SSSP), used to extrapolate to future
architectures~\cite{8049025}, or used to evaluate vendor-proposed microarchitecture features~\cite{8049003}.
Moreover, these proxies needed to be further scaled down in complexity and runtime so that they could be used
with cycle-accurate processor simulators~\cite{9410320}, such as gem5.

Overall, many of the proxy applications listed above have been widely used for procurement decisions, vendor
interactions (where full applications are too complex or classified), evaluation of programming models
and testbeds, and supercomputing research all the way from transistor to full system scales (see various related
scientific papers and presentations~\cite{domke_hyperx_2019,osti_1645768,osti_1806777,osti_1763624,osti_1608914,osti_1477829,osti_1617284,9229635,9460517,richards-glenski-best-practices}). 
However, the listed proxies are not the only ones available. For example, the Standard Performance Evaluation
Corporation (SPEC)~\cite{spec_spec_nodate} and the International Open Benchmark Council (BenchCouncil)~\cite{benchcouncil}
offer a wide range of proxy applications for single-/multi-threaded CPUs, distributed systems, accelerators, artificial
intelligence, etc., which are used by HPC centers, vendors, and academia. Furthermore, the increasing need to perform
big data processing and machine learning on supercomputers and data centers also proliferated the design of
proxies for these areas, e.g. Intel’s HiBench~\cite{5452747}, DeathStarBench~\cite{10.1145/3297858.3304013},
Baidu's DeepBench~\cite{baidu_inc._DeepBench_2017}, and MLCommons' MLPerf~\cite{MLSYS2020_02522a2b}; and large data center
operators, such as Google, open-sourced their (meta-)frameworks for procurement evaluations~\cite{google_perfkit}.
And all these benchmarks are in addition to the more traditional, yet still relevant, peak performance benchmarks, such as High Performance Linpack (HPL)~\cite{dongarra_linpack_1988} and Conjugate Gradient (HPCG)~\cite{dongarra_hpcg_2015},
stream~\cite{deakin_gpu-stream_2016}, and Intel's MPI Benchmarks~\cite{intel_corporation_intel_2018}, to name just a few.

\mbox{}\newline\noindent\textbf{Takeaway messages of Section 2:}
\begin{itemize}
    \item Proxy applications were indispensable for the success of co-design efforts for the pre-exascale and early exascale supercomputers.
    \item Future workload demands, and workload changes due to algorithmic changes, cannot be captured by proxy apps.
    \item Supercomputing and data centers are ``drowning'' in, mostly overlapping, benchmarks and proxy applications.
\end{itemize}

\section{Lessons-Learned from the Past Decade}
An in-depth, successful co-design of modern supercomputers without the assistance of proxy applications and workload
analysis would have been unlikely, but the current state of our co-design ``toolbox'' contains some remediable
and some fundamental flaws which we (HPC centers) and vendors uncovered while working with the benchmarks and tools.

\subsection{The Evolution of Proxy-applications}
The community did not waste effort in trying different approaches for designing proxy-applications and benchmarks to be used in procurement.
In the build up to the petascale era in the early 2000s, the community recognized the necessity for moving away from relying on workload model 
benchmarks given how working with such models is complex for small companies and also the lack of utility for assessing heterogeneity.
The HPC challenge benchmark suite in 2005~\cite{luszczek:2005:introduction} and later the HPCS program by DARPA paved way for the first transition
to small and simple benchmarks~\cite{dongarra:2008:darpas}. The main point for the HPCS program was to provide benchmarks that bound the
performance of many real applications as a function of memory access characteristics and then use a correlation matrix to map the properties of the
parameterized benchmarks back to applications~\cite{weinberg:2005:quantifying}. However, HPCS was considered to not have enough fidelity for
procurement activities. That is mainly due to: a) the inability to relate the numbers reported by the benchmarks to real workloads, and
b) the rigidity of the benchmarks that hindered efforts for emphasizing memory hierarchy and locality or exploring heterogeneity.

The lessons from HPCS led to the next transition: mini-applications and proxy-applications. After recognizing the limitations of the benchmark approach
of HPCS, the community moved to proxy-applications in the Exascale Computing Project (ECP)~\cite{osti_1477829,exascale_computing_project_ecp_2018}.
Proxy-applications were carefully selected to characterize a broad-spectrum of applications. And while they have been used in the process of procuring the current generation of system, they too have their limitations, as will be discussed in following sections.    

\subsection{Implementation Biases}
Any implementation of a benchmark, no matter the size, entails multiple implicit biases towards programming language, data layout,
and parallelization approach, but also towards the underlying algorithm which is used to solve a problem. Since many of
the ECP, Fiber, and SPEC proxy application are scaled-down versions of existing HPC workloads, they are still predominantly
implemented in Fortran and C, and have been tuned over the years, and sometimes decades, for cache-based CPU architectures.
This over-commitment of co-design in one area can lead to lack of attention in other areas, causing, for example, under-performing C++
applications with certain compilers. Similarly, the extensive focus on GPGPU-based supercomputing resulted in highly tuned
CUDA (and ``legacy'' CPU) proxy applications from the ECP side with memory/data layouts optimized for those architectures.
These codes and layouts are not easily transferable to OneAPI/SYCL which will be used in ANL's Aurora exascale
supercomputer, and hence more engineering and time was required to port the codes to another programming paradigm.
Another bias we uncovered is the reliance of a climate simulation proxy on compiler-based auto-parallelization of inner loops,
which, when undocumented, can cause major issues for anyone tuning/porting the code for other architectures. The existence of proxy
applications creates an even higher barrier of entry into the HPC field for emerging and alternative processor designs, such as quantum,
dataflow, or photonics-based, because these solutions require a change of the algorithms. For example, traditional implementations
solve NP-hard problems (e.g. quantum chemistry) using iterative methods or heuristics, while a quantum computer will
require a quantum algorithm, written in a new programming model like QUBO~\cite{butko2020tiger}, but consequently could outperform traditional CPUs by many orders of magnitude.

\subsection{Complexity Trap and Performance Portability Myth}
Unfortunately, and despite best efforts by the HPC community, achieving performance portability across slightly or widely
varying architectures is still an unobtainable goal. Modern features in OpenMP or alternative programming paradigms, such as
OpenCL, demonstrate that applications can be written (often without separate code paths for different architectures) in a way to
easily migrate between different CPUs and GPUs and vendors while preserving correctness, however usually such migration
results in severe performance drops~\cite{10.1145/3404397.3404441,inbookopenmp,PENNYCOOK20131439}.
Hence, manual code refactoring is still required, to change algorithms, data layouts, parallelization strategies, etc.,
to fully utilize any given architecture. However, domain scientists estimate the cost of such refactoring efforts
(for example, to target FPGAs) of upwards of 11 million dollar per 100.000 lines of code~\cite{hyperion_nasa}.
Such cost for a single proxy is obviously unjustifiable for vendors and HPC centers during the early phases of the co-design, i.e., while
exploring diverging architecture options, especially when considering the growing number of proxies and code complexity
thereof. Therefore, the focus on proxies of large/legacy scientific codes, which had been tuned for previous architectures, traps
the co-design participants into considering only similar architectures.

\subsection{Insufficient Input Configurations, Testing, and Documentation}
We noticed additional short-comings while working with many of these proxy applications in multiple research and co-design projects.
For example, some applications rely on third-party, sometimes closed-source, libraries or inputs which were inaccessible
because the documented hyperlinks were outdated and content had been moved. Working with different compilers or even compiler
version proved challenging, since many of the scaled-down workloads have not been tested across a range of compilers, and
changes in the environment exposed implementation errors or numerical instabilities. Similarly, the compatibility with external
performance profilers and analyzing tools, such as Intel Advisor, Score-P, TAU, etc., is not always guaranteed, yet should be part of the focus~\cite{richards-glenski-best-practices}.
Furthermore, the documentation of the proxy applications often ranges in quality, even within certain sets of benchmarks from the
same institution, with lack of information such as: (a) which parts of the code are performance-critical, (b) is bit-reproducibility
required, (c) what is/are the implemented algorithms and are there alternatives, and (d) which system characteristics does this input evaluate?, to name just a few.
Anecdotal evidence indicates that vendors can optimize the wrong aspects when there is a lack of information and inaccurate
representation of the proxy w.r.t the real application, and for example change the input from an unstructured mesh to a structured mesh to
gain performance, because a proxy used a structured mesh generator to transform it into unstructured mesh, while the HPC center intended
to use it as ``unstructured mesh''-proxy.
Such inconsistencies could have been avoided if all participants were aware of existing guidelines,
for example the ones outlined by members of the Advanced Simulation and Computing (ASC) Program~\cite{osti_1055856},
and would have followed them when designing the proxy.
Last but not least, many proxy applications lack variability in input selection, i.e. they lack
strong or weak scaling tests, have fixed requirements for memory, OpenMP threads, MPI ranks, etc., or lack meaningful inputs
which are usable with slow, high-fidelity simulators. These shortcomings can severely impede the co-design approach, however most
of them are preventable with adequate funding, community effort, and predetermined guidelines for code/input/documentation quality.

\subsection{Lack of Efficiency Reporting}
It is important for performance engineering practitioners and end users to understand the efficiency a given benchmark could
achieve on a given hardware target. Efficiency in this context can be viewed as \emph{how much performance was achieved, in
comparison to the peak theoretical performance}. All the benchmark suites listed in earlier section do not report efficiency.
Explicitly reporting efficiency is more increasingly important as we head into an era of more complexity in systems; without
keeping all stakeholders aware of the efficiency considerations, we run the risk of designing complex systems that are poorly
utilized and/or hard to program. Additionally, reporting the efficiency in benchmarks further helps users to adjust their
expectations when porting their codes to new systems.  

\subsection{Vendor Feedback}
HPC vendors and system integrators, such as AMD and HPE/Cray, also expressed their views on proxy applications publicly~\cite{amd_codesign,amd_codesign2,richards-glenski-best-practices}
and privately. Proxies create considerable labor costs when porting to new architectures, requires expert knowledge of the scientific
problem or extensive collaboration with the domain scientists, and can inaccurately represent computational or data movement bottlenecks
if the proxy diverges too far from the real workload due to the reduction in scale or complexity. AMD recommends to change the focus
from proxies to extensive data collection on instrumented production hard- and software and utilization of machine learning to capture
workload behavior. However, this results in the same implementation bias which we are trying to overcome, with the additional caveat of having
an adversarial impact on the performance of production system when the necessary sampling frequency is too high.
Meanwhile, the feedback from HPE/Cray (echoing some of AMD's feedback) additionally stresses the focus on lifting many restrictions
in the benchmarks, such as the amount of nodes/cores/etc, and the requirements for bit-identical reproducibility which would
hamper the introduction of better architectures. Benchmarks should be meaningful for the workloads at the site, properly documented,
and allow for optimizations, and expectations (w.r.t. the vendor's labor cost) should be scaled according to the overall RFP budget.

\subsection{Current Efforts to Overcome the Shortcomings}
With the focus of centers shifting to prepare their priority workloads for delivered systems, the work on proxy-apps has slowed. Few efforts, e.g. filtering duplicated proxies via $\cos\theta$-metric~\cite{richards-glenski-best-practices}, are still ongoing, but no coordinated and noticeable activity is dedicated towards resolving the proxy-application limitations we mentioned.
In the reminder of this article we layout our vision for the future: evolving beyond proxy-applications.

\mbox{}\newline\noindent\textbf{Takeaway messages of Section 3:}
\begin{itemize}
    \item Proxy applications, by pre-fixing the \textit{question->...->language} parameters, are impeding the adoption of novel execution engines.
    \item Porting complex proxies becomes prohibitively expensive, due to necessary changes of data layout, language, and algorithms,
    and further requires deep knowledge of the underlying mathematical problem.
    \item Without transparency in porting the proxies, the eventual users of the new systems are not well-informed of the efforts entailed to port their codes. 
    \item The compatibility of proxies and other co-design tools, such as profilers, simulators, etc., is not implicit.
\end{itemize}

\section{The Co-Design Toolbox of the Future}
Since neither micro-benchmarks nor application proxies seem to be the right fit for the required co-design in the next decade, one being
too detached from reality the other too static and backward looking, we require something in-between. We envision a form of highly-parameterizable,
easily-amendable, Motifs-like representations of algorithms or kernels. Such complex ``operations''
could be anything from Fast Fourier Transform (FFT) or sparse matrix-multiplication over data-structure padding operations for stencils to
re-meshing operations in load-balanced solvers. For their shape-shifting nature, we call them \textit{Octopodes} instead of
scientific Motifs~\cite{Asa:EECS-2006-183} (e.g. NAS Parallel Benchmarks~\cite{bailey_nas_1995}), since they are more abstract
and meant to be closer to algorithms (supported by one/many reference implementations) than fixed proxy-implementation.
The utility of \textit{Octopodes} and how they fit into the remaining co-design toolchain is subject of the following subsections.

\subsection{Representative \textit{Octopodes} for Co-Design}
The goal is to capture the algorithms or complex operations instead of the implementation, which make up modern workloads, and which can be optimized, transformed, or replaced.
There are precedents for such highly-parameterizable kernels which can be rapidly generated, for example SPIRAL~\cite{Franchetti:18} can be used to
synthesize and autotune high-level specifications of mathematical algorithms, such as DFTs~\cite{Popovici:20}, for specific hardware.
Assuming, we have such kernel- or operator-generation capabilities for a wide range of typical scientific problems, then it could be possible to
use machine learning for:
\begin{enumerate}
    \item[(a)] automatically identifying compute phases or regions-of-interest across all scientific codes and projects running on a given supercomputer,
    \item[(b)] determining the right parameterization to correlate an \textit{Octopode} to a real application region, and
    \item[(c)] finding similarities between applications (for example, by using the $\cos\theta$-metric~\cite{richards-glenski-best-practices});
\end{enumerate}
such that hardware and software tuning can be performed more efficiently for a given problem class instead of ten different (proxy-)
applications which all exhibit similar behavior.

As in the case of SPIRAL, such high-level specifications and parameterizations of interest, can be used to convey the mathematical problem
and underlying compute pattern to a co-design team or hardware vendor in a more descriptive manner than a million lines of source code and 
scientific papers.
The co-design team would be able to experiment with data layout changes, hardware options, and even algorithmic alternatives for the same problem.
For HPC procurement, the parameterization could be restricted and/or any changes, such as numerical precision or data layouts, would need to be
transparently documented and measured for specific input sizes, for example: \textit{``algorithm X was replaced with Monte Carlo-based algorithm Y
and pre-/post-algorithmic layout changes require 2\% and 5\% overhead, respectively, but it yields a benefit of 7x speedup over baseline''}.
This allows for the required flexibility to explore and incorporate alternative architectures in a more efficient way.

What we mean by ``high-level specifications and parameterizations'' can be more easily understood when looking at another potential \textit{Octopode},
namely matrix-multiplications or in short \emph{matmul}, which everyone in the HPC community is familiar with.
Instead of just benchmarking double-precision dgemm with HPL, a single matmul-\textit{Octopode} would support various input shapes (such as squared,
rectangular, and tall/skinny) and numerical precisions (i.e., from quadruple precision (fp128) as used in some quantum simulations, all the way
down to low-precision such as bfloat16 and the like), and batched and non-batched execution. Furthermore, the matmul-\textit{Octopode} shall not
only cover dense matrix-matrix operations, but also matrix-vector, and sparse matrices. Many of the sparse matrix formats exploit local memory
to some degree to achieve computational performance, but such blocking needs to be supported by the input matrix, and hence a simple randomly
generated sparse matrix is unsuitable as general input representation, and should only be used as one of many. Other realistic sparse inputs can
be sampled from public repositories~\cite{Kolodziej2019,Boisvert1997} or various existing HPC workloads (e.g. resulting from structured meshes) and
DL problems, which often result in highly regular and blocked sparsity. The required parameterizations can be expressed via template metaprogramming
in C++, for example, without too much engineering overhead, which was not available or mainstream during the HPCS program.

Other \textit{Octopodes} can be either derived from existing works, such as the Apex-MAP~\cite{weinberg:2005:quantifying} or Siena~\cite{peng:2018:siena},
or can be implemented from scratch. The former, Apex-MAP, is a synthetic benchmark that stresses a machine’s memory subsystem according to parameterized degrees of spatial
and temporal locality, and it could be retrofitted to generate many realistic access patterns to feed machine learning models. Siena, which is designed to generate
load/store and compute patterns to quickly explore diverse memory architectures, on the other hand, can be used to tune the ML models and prevent
overfitting for a certain architecture or testbed, to improve the accuracy of identifying an \textit{Octopode} in application regions, see points (a) and (b) above.

\subsection{Emerging Workloads or Science Domains and End-to-End Workflows}
The \textit{Octopodes} do not only exist in traditional HPC applications, but also in other big data workloads. A prominent example are Deep Neural Networks (DNNs)
which contain tens or hundreds of layers. Each of those layers, or even multiple layers fused together, could be represented by \textit{Octopodes}. For example,
convolutions layers, depending on implementation, can be similar to matrix multiplication, FFT, or to stencil operations. A ``generic'' \textit{Octopode} for
stencil operations can be parameterized to identify and match these DNN layer operations, or another ``generic'' \textit{Octopode} for high-dimensional array
transformations (e.g., matrix transpose) can capture tensor layout changes, which are required for the data flow between (fused) layers, if parameterized correctly.
Similarly, a flexible \textit{Octopode} for scatter and/or gather operations could reflect the data flow in map-reduce workloads or represent interactions with
the parallel filesystem, independent of (but parameterizable for) the actual implementation.
This also holds true of other existing and future workloads, since they are all constructed from a sequence of algorithms or complex operations on data
structures induced by inputs which can be artificially in-/decreased to match current or expected future configurations, e.g. inferred from the desired
resolution and components of a climate simulation.

\subsection{The future Co-Design Cycle with \textit{Octopodes}}
Our vision for a modernized co-design approach requires even tighter collaboration between the HPC users and the co-design teams, and allows more flexibility for the vendors.
The first step requires the users and co-design teams to analyze the dominant HPC workloads (w.r.t. the consumed node-hours). This process consists of: (a) profiling the execution,
(b) identifying regions-of-interest, (c) collecting performance-relevant data such as execution time and hardware counters, and (d) categorizing the regions into algorithms
and complex operations. In the past, (a-c) were commonly done by users, but (d) is necessary as well for a holistic view, for machine learning, and for an improved co-design.
Such information should be accessible across multiple HPC centers and countries and contain enough data to enable sophisticated ML techniques. The goal is the
identification capability of regions-of-interest (or potential \textit{Octopodes}), preferably in an automatic way. These ROI can then be mapped to existing
\textit{Octopodes}, or they will have to be transformed by users, co-design teams, and vendors into novel \textit{Octopodes}, which correlate to true HPC and
data center workloads, if parameterized appropriately. Together with a curated list of micro-benchmarks
and slimmed-down number of (non-overlapping) proxy-application, these \textit{Octopodes} build the targets for the co-design, where micro-benchmarks are used to specify
and test the necessary peak performance of a system, \textit{Octopodes} are to be used in conjunction with other co-design tools (e.g. architecture simulators) to select the
best hardware for existing and predicted/future workload by the co-design team and the hardware vendor. 
To demonstrate the hardware capabilities, the vendors shall be allowed more tuning freedom for the \textit{Octopodes}, i.e., changes of algorithm, implementation,
integer/floating-point precision, data layout, etc., as long as the intended result remains the same, the changes are properly communicated, and not only the algorithm is benchmarked (but also the
necessary time for the pre/post-execution transformation of the data). 
These smaller \textit{Octopodes}, in comparison to proxies, are also more amenable to automated performance tuning, or can be used as design targets in AI/ML-driven
architecture generation, as it has been demonstrated recently~\cite{Mirhoseini2021,aart-de-geus}.
The knowledge for mapping an \textit{Octopode} to a given hardware can then be used in a very limited set of proxy-applications to serve as a benchmark suite
for acceptance testing and to serve as a demonstrator for the users on how to change/tune their codes for the next system.

The crucial aspects for this design cycle to succeed are better tools, extensive automation in workload analysis and architecture modelling and evaluation, and increased
bidirectional knowledge transfer between users, system operators, co-designers, and hardware/software vendors. Furthermore, \textit{Octopodes}, as well as micro benchmarks and
proxies, need to be appropriately documented, e.g. which results are considered canonically correct when comparing algorithms or when comparing von Neumann architectures and
quantum computers, and what are the options and the rules for re-implementing a given algorithm or complex operation.
Finally, all (reference-) implementations of \textit{Octopodes} should, in addition to reporting architecture-independent performance metrics (e.g. work/time), report the efficiency
for the implementation when run on a target hardware.
The roofline~\cite{10.1145/1498765.1498785} model can be used for this purpose. While we acknowledge the limitations and over-simplification of the roofline model,
it is nonetheless simple to produce, commonly used in the HPC community, and suffices as a first order approximation for efficiency from which an observer can make
fast assessments.

\mbox{}\newline\noindent\textbf{Takeaway messages of Section 4:}
\begin{itemize}
    \item The \textit{Octopodes} are mutable, high-level, algorithmic specifications and problem parameterizations which can represent compute phases, such as
    complex operations or entire algorithms, within larger scientific HPC workloads.
    \item Future \textit{Octopodes} should not serve as an ultimate tuning target; they should foster a shared understanding among all co-design participants of what it means to be ``good'' for a given problem.
    \item We see a growing need for machine learning and other automation tools for specifying, generating, and identifying \textit{Octopodes} which can aid in the co-design cycle
    for future supercomputers.
    \item Users benefit from the developed tools, since these tools similarly assist in code refactoring as they assist in the co-design.
\end{itemize}

\section{Outlook and Summary}
The use of proxy applications expanded and improved the co-design capability of modern supercomputers, but we believe that current hardware trends and software complexities
require a new set of tools for the co-design of post-exascale supercomputers and federated HPC/data centers to better capture, analyze, and model existing and
future workload demands. To open the floor for future, community-wide discussions, we have outlined the state-of-the-art and its shortcomings, and propose an alternative, hopefully better suited set of highly-parameterizable,
easily-amendable, Motifs-like problem representations which we call \textit{Octopodes}. These algorithms or complex operations shall not replace proxy applications
entirely but supersede them as the primary target in the co-design process. Octopodes will be the common language between HPC users, system operators, co-designers,
and vendors to describe the to-be-solved scientific challenges, what needs to be computed, and how it can be computed, in an abstract way.
This approach allows
for more flexibility in the hardware/software design and selection to match the users needs with the best architecture, instead of fine-tuning legacy architectures
to legacy implementations. In this position paper, we demonstrate our idea of future \textit{Octopodes} by multiple examples, such as the highly versatile matrix multiplication
and auto-generated DFT kernels, and how they, together with machine learning-assisted tools, will help system architects and HPC users. We expect that our conception of a community-driven and
well-curated set of \textit{Octopodes} is able to improve the overall co-design cycle, while also being able to alleviate the increased complexity and labor cost associated
with modern proxy applications.

\bibliography{main_refs}

\end{document}